\documentclass{elsart3}

 \usepackage{graphicx}
\usepackage{epsfig}

\usepackage{amssymb}

\usepackage{bm}

\begin{document}

\begin{frontmatter}



\title{Five Years of Tracking Heavy Ion Collisions at RHIC}


\author{Achim Franz}

\address{Brookhaven National Laboratory, Physics Department, Bld 510C, Upton, NY 11733-5000, USA}

\begin{abstract}
Five years have passed since the first collisions of Au nuclei at the Relativistic Heavy Ion Collider (RHIC) at 
Brookhaven National Laboratory (BNL) on Long Island. With nucleon-nucleon center-of-mass energies of up to $\sqrt{s_{NN}}=200~GeV$ RHIC provides 
the highest energy heavy ion collisions at any existing collider. 
To study the dynamics of nuclear matter at extreme temperatures and pressures  
hundreds of produced particles need to be tracked and identified, 
which provides a sizable challenge to the four experiments. This article tries to summarize these first years of 
RHIC operation from the detector point of view and give a glimpse at the future of the accelerator and its experiments. 
\end{abstract}

\begin{keyword}
RHIC \sep heavy ion \sep collider \sep tracking 
\PACS 
\end{keyword}
\end{frontmatter}

\section{Introduction}
\label{Sec:Intro}

The history of RHIC and its experiments goes back to the early 1980's and is summarized e.g. in \cite{baum2001}. 
After an initial engineering run in 1999 cool-down of the two 3.8 km long cryostats containing about 1000 
super-conducting magnets started in February 2000 and operation with gold beams, delivered by the injection 
chain consisting of Tandem, Booster, and AGS, started in April. 
First collisions in all four experiments  were observed  during June 2000 and a luminosity of 
$\mathcal{L}=2\times10^{25} cm^{-2} s^{-1}$ (10\% of design) has been achieved.

Since then a total of five runs have produced billions of events, peta-bytes (PB) of data, 0.5~PB for the 2004/2005 PHENIX dataset alone, 
at various beam energies and collision species of up to twice the design luminosity, see Table~\ref{Tab:RHICLumi}, 
resulting in 122 articles in peer-reviewed journals and 103 PhD thesis to-date. 
A comprehensive summary of the physics results from the four
experiments covering the first three years is available in \cite{whitepaper}. 

\section{The RHIC complex}
\label{Sec:RHIC}

\begin{figure*}[htbp]
\begin{center}
\epsfig{file=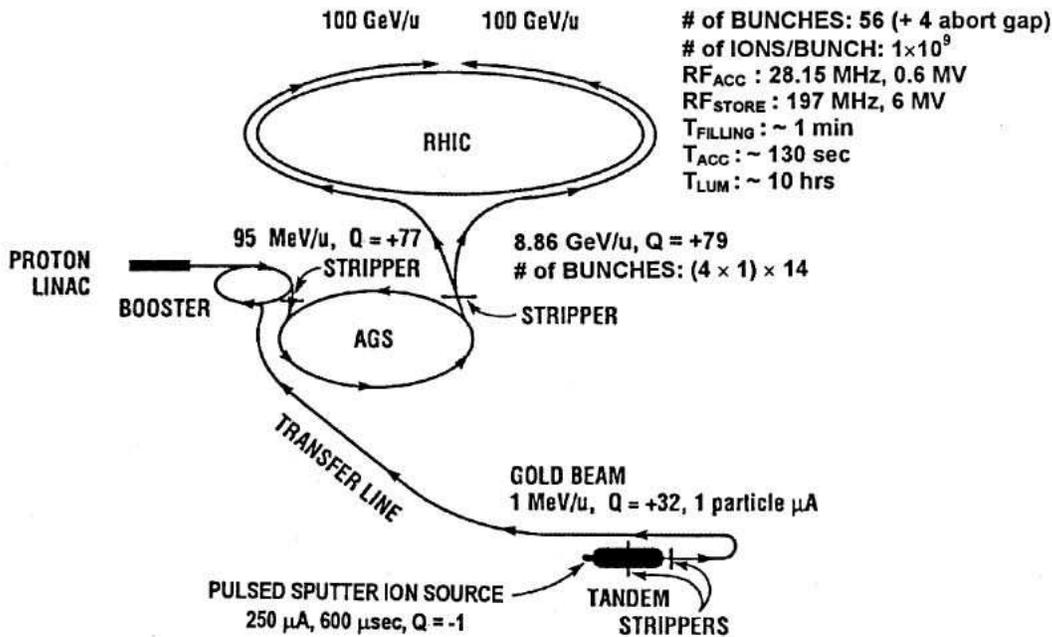,scale=0.5}
\caption{The RHIC complex for Au collisions \cite{rhicnim01}.}
\label{Fig:rhiclayout}
\end{center}
\end{figure*}

Figure \ref{Fig:rhiclayout} shows the layout of RHIC for Au running \cite{rhicnim01}. Ions from the Tandem are accelerated 
in the Booster then collected and bunched in the AGS and injected into the two separate 3.8 km long RHIC rings. 
Protons for the polarized proton program are injected from a Linac, middle-left on Fig.\ref{Fig:rhiclayout}.
A complete overview of the RHIC accelerator complex can be found in \cite{rhicnim}.  

Already in the second year of operation RHIC achieved design luminosity, but with the drawback of high backgrounds 
and short times in store mainly due 
to intra beam scattering. A better vacuum in the beam pipes, especially the warm sections, was achieved by bake-out and NEG coating. 
The use of Storage RF and collimators have further improved the beam quality.
However the experiments were still suffering from the background levels, specially during the beginning of each store. 
This made the operation of some wire chambers impossible, e.g. the muon chambers installed in PHENIX suffered from rates 100 times higher than 
was estimated from collisions and expected beam background. 
Therefore, extensive shielding had to be installed in the tunnel close to the interaction regions. Measurements 
close to the beam-pipe at the PHOBOS experiment indicate an average dose of 1 kRad/year \cite{pakprivate}.

Details about the 
RHIC performance of the last years can be found in Table~\ref{Tab:RHICLumi}, the RHIC run summary web page \cite{rhiclumi}, 
and in the proceedings of recent (European) Particle Accelerator Conferences (PAC, EPAC) \cite{EPAC04,PAC05}.

\begin{small}
\begin{table*}[htdp]
\caption{RHIC operating modes and total integrated luminosity delivered to the experiments \cite{rhiclumi}}
\begin{center}
\begin{tabular}{| l || r@{ - }l | r@{.}l | r@{.}l | c |} \hline

~    & \multicolumn{2}{c|}{~}               & \multicolumn{2}{c|}{particle} & \multicolumn{2}{c|}{total}           & average      \\ 
~    & \multicolumn{2}{c|}{~}               & \multicolumn{2}{c|}{energy}  & \multicolumn{2}{c|}{delivered}  & store       \\
runs & \multicolumn{2}{c|}{species} & \multicolumn{2}{c|}{[GeV/n]} & \multicolumn{2}{c|}{luminosity} & polarization \\ \hline

Run-1 2000    & $Au^{79+}$&$Au^{79+}$ &  27&9 &  $< 0$&$001 \mu b^{-1}$ &     \\ \cline{2-7}

              & $Au^{79+}$&$Au^{79+}$ &  65&2 &   $20$&$0 \mu b^{-1}$     &     \\ \hline\hline

Run-2 2001/02 & $Au^{79+}$&$Au^{79+}$ & 100&0 &  $258$&$0 \mu b^{-1}$     &     \\ \cline{2-7}

              & $Au^{79+}$&$Au^{79+}$ &   9&8 &    $0$&$4 \mu b^{-1}$   &     \\ \cline{2-8}

              &      $p^{+}$&$p^{+}$  & 100&0 &    $1$&$4 pb^{-1}$      & 14\% \\ \hline\hline

Run-3 2002/03 & $d^{+}$&$Au^{79+}$    & 100&0 &   $73$&$0 nb^{-1}$        &     \\ \cline{2-8}

              &      $p^{+}$&$p^{+}$  & 100&0 &    $5$&$5 pb^{-1}$      & 34\% \\ \hline\hline

Run-4 2003/04 & $Au^{79+}$&$Au^{79+}$ & 100&0 & $3740$&$0 \mu b^{-1}$     &    \\ \cline{2-7}

              & $Au^{79+}$&$Au^{79+}$ &  31&2 &   $67$&$0 \mu b^{-1}$     &      \\ \cline{2-7}

              &      $p^{+}$&$p^{+}$  & 100&0 &    $7$&$1 pb^{-1}$      & 46\% \\ \hline\hline

Run-5 2004/05 & $Cu^{29+}$&$Cu^{29+}$ & 100&0 &   $42$&$1 nb^{-1}$      &      \\ \cline{2-7}

              & $Cu^{29+}$&$Cu^{29+}$ &  31&2 &    $1$&$5 nb^{-1}$      &      \\ \cline{2-7}

              & $Cu^{29+}$&$Cu^{29+}$ &  11&2 &    $0$&$02 nb^{-1}$     &      \\ \cline{2-8}

              &      $p^{+}$&$p^{+}$  & 100&0 &   $29$&$5 pb^{-1}$      & 46\% \\ \cline{2-8}

              &      $p^{+}$&$p^{+}$  & 204&9 &    $0$&$1 pb^{-1}$      & 30\% \\ \hline\hline

\end{tabular}
\end{center}
\label{Tab:RHICLumi}
\end{table*}
\end{small}

\section{The Experiments}
\label{Sec:experiments}

Before data from the RHIC experiments were available, models predicted the charged particle multiplicity at mid-rapidity 
to be between 600 - 1400 particles. Consequently the experiments were designed to cope with the highest 
multiplicities predicted. 

\begin{figure*}[htbp]
\begin{center}
\epsfig{file=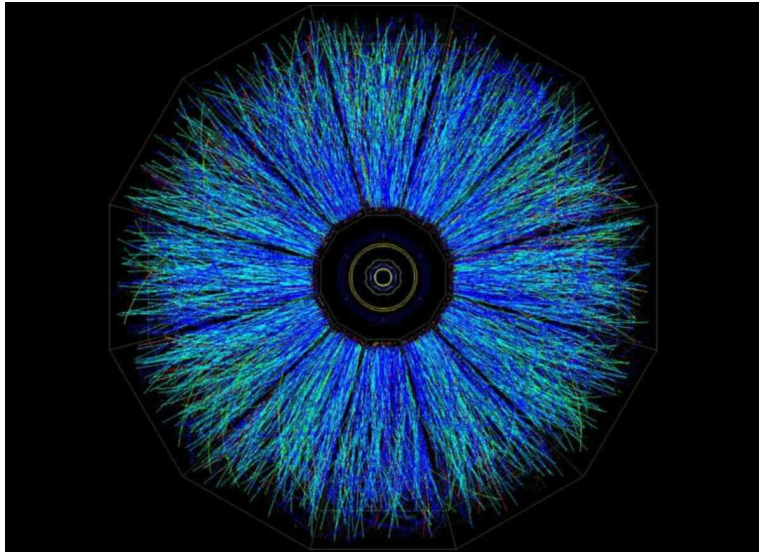,scale=1.0}
\caption{Beam’s eye view of a central event in the STAR Time Projection Chamber. This
event was drawn by the STAR level-3 online display.}
\label{Fig:starevent}
\end{center}
\end{figure*}

Comparing the first results, see Fig. \ref{Fig:starevent}, presented at the Quark Matter Conference 2001 with model predictions at that time \cite{eskola2001} 
revealed that the measured multiplicity was at the low end of the spectrum. This not only aided the RHIC experiments 
but made it feasible for the LHC p-p experiments to participate in the heavy ion program, where up to 8000 charged particles are expected.

\begin{figure*}[htbp]
\begin{center}
\epsfig{file=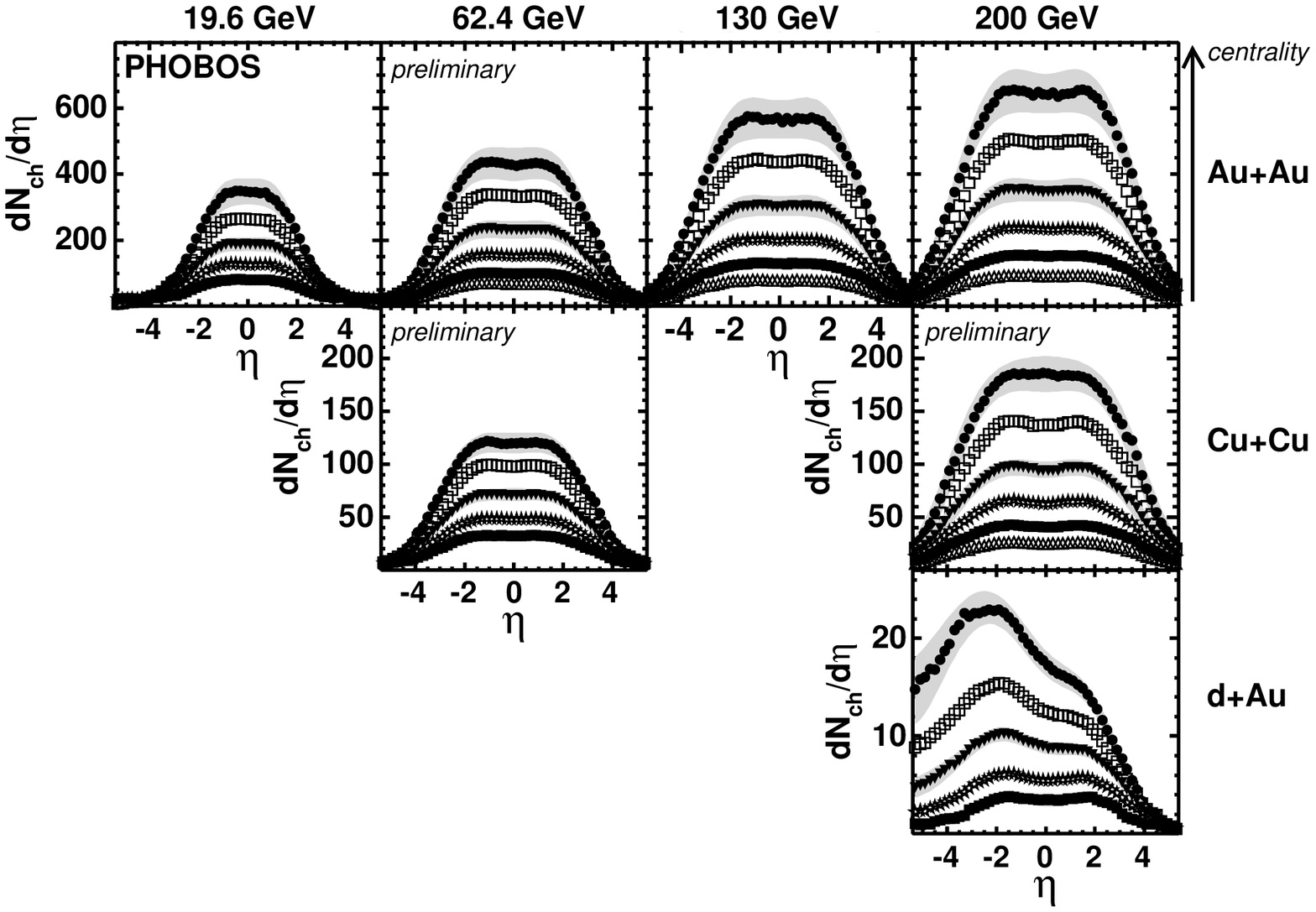,scale=0.65}
\caption{Charged hadron pseudorapidity distributions for different collisions centralities (see text) for
Au-Au collisions at $\sqrt{s_{NN}}$ = 19.6, 62.4, 130 and 200~GeV (top row from left to right),
Cu-Cu at 62.4 and 200~GeV (middle row, preliminary) and d-Au at 200~GeV (bottom row) \cite{Roland0510042}.}
\label{Fig:phobosmult}
\end{center}
\end{figure*}

Figure \ref{Fig:phobosmult} shows a set of charged particle multiplicity measurements for three different collision systems, 
four energies and multiple collision centralities \cite{Roland0510042}.
These RHIC data combined with SPS measurements extrapolate to a LHC Pb-Pb multiplicity of around 1000~-~2000 at mid-rapidity. 
How the LHC experiment CMS is preparing for the heavy ion program can be found in \cite{roland2005}.

Heavy ions are extended objects and their collisions is not always head-on, the degree of geometrical overlap, or centrality, is described by the impact parameter, b, 
or perpendicular distance to the closest approach if the ions were undeflected. 
Zero degree calorimeters (ZDC) \cite{zdc01,zdc02}
have been installed in all experiments
to provide a collision and vertex location trigger as well as a measure of the collision centrality by the energy deposited from the 
forward going spectator neutrons. 
A complete geometrical overlap of the ions, $b\approx0$, is called a central or large centrality event
resulting in e.g. a high charged particle multiplicity and very few or none forward going neutrons.
The ZDCs also provide feedback to the accelerator operators
for steering and to estimate the delivered luminosity. 
In addition to the ZDCs all experiments have collision triggers installed, usually quartz Cherenkov or Scintillator detectors, 
which provide a trigger for the vertex location, in the cm range, and a start time for drift or time-of-flight (TOF) measurements. 

Four experiments are participating in the heavy ion program at RHIC:\\

{\bf B}road {\bf Ra}nge {\bf H}adron {\bf M}agnetic {\bf S}pectrometer 
(BRAHMS) \cite{BRAHMS01}. \\ 
The collaboration consists of 56 physicists from 12 institutions in 5 countries. 
The experiment is a small acceptance spectrometer consisting of a set of global detectors for event characterization and two spectrometer arms 
designed to identify hadrons over a wide range in rapidity and momentum. 
The two arms are used to cover the 
mid-rapidity range (MRS 6.5~msr) where the particle momentum is in the few hundred MeV/c range, and a movable ($30<\vartheta<90$) forward arm (FS 0.8~msr) 
which can identify particles up to 30~GeV/c. 

The particle identification (PID)
is performed with standard spectrometer components like Time-of-Flight (TOF), threshold and ring-imaging Cherenkov detectors and tracking devices 
like drift (DC) and time-projection (TPC) chambers. 
The TPCs are of standard design with 21.8~cm drift (229 V/cm) using $Ar/CO_{2}$ (90:10) and STAR front-end-electronics (FEE). 
The DCs have 10 layers with x, y, and $\pm18^{\circ}$ wire planes
using a $Ar/C_{4}H_{10}$ (67:33) gas mixture with a $9^{\circ}C$ alcohol bubbler. 
Due to the low track density of 0.01 to 0.1 $cm^{-2}$ a simple 'follow-your-nose' algorithm is used for tracking.  
This is a simple approach in which a 
hypothesis or previously determined points are used to predict where the next measured point on the track 
should be.

The compact size of the detectors are ideal for construction and maintenance. In one of the earlier runs a field wire in the TPC broke, 
but being $< 0.2m^{3}$ in size they can easily be extracted and repaired in the lab. None of the tracking chambers as shown any aging effects. \\

{\bf P}ioneering {\bf H}igh {\bf E}nergy {\bf I}nteraction E{\bf x}periment (PHENIX) \cite{PHENIX01}. \\
The collaboration consists of 498 physicists from 62 institutions in 13 countries. 
The experiments was 
designed to measure mainly electrons, muons, and photons, with a good hadron PID at mid-rapidity. 
It has two distinct detector sections. At mid-rapidity two arms with 
tracking, PID and calorimetry measures hadrons, leptons and photons, whereas in the forwards directions  muons are identified and tracked. 

The muon tracking system \cite{akikawa} 
uses three layers of cathode strip chambers inside the muon magnets and 6 layers of Iarocci drift tubes interleaved 
with several inches of steel. These chambers suffered the most from high beam background but no aging effects were observed. 
Because they are located inside the magnetic field the strip chambers include an optical alignment system \cite{murata}.

The mid-rapidity arms \cite{adcoxcentral} feature, apart from PID detectors, three types of wire chambers, 
a jet-type drift- (DC), 2(3) layers of pad- and 6 layers of 
time-expansion/transition radiation- (TEC/TRD) chambers. 
Early on HV instabilities required the addition of an alcohol bubbler ($0^{\circ}C$) 
to the DC gas of $Ar/C_{2}H_{6}$ (50:50). The TEC/TRD ($Xe/He/CH_{4}$ 45:45:10) shows a slight decrease in gain over the past few years
which could be a sign of material induced aging or deposits on the field wires. 

In 2005 the multiplicity-vertex detector (MVD) \cite{allen}, a silicon strip (barrel) and pad (end-cap) detector, was removed from the setup because noise induced fluctuations in the pedestal values 
made reliable physics measurements impossible. It will be replaced in the coming future by a multi-layer vertex tracking detector \cite{heuser}.\\

PHOBOS \cite{PHOBOS01}.
The collaboration consists of 68 physicists from 8 institutions in 3 countries.
The name of the experiment is not an abbreviation, it was chosen from the name of one of the moons of Mars, which was the first name of the experiment. 

\begin{figure*}[htbp]
\begin{center}
\epsfig{file=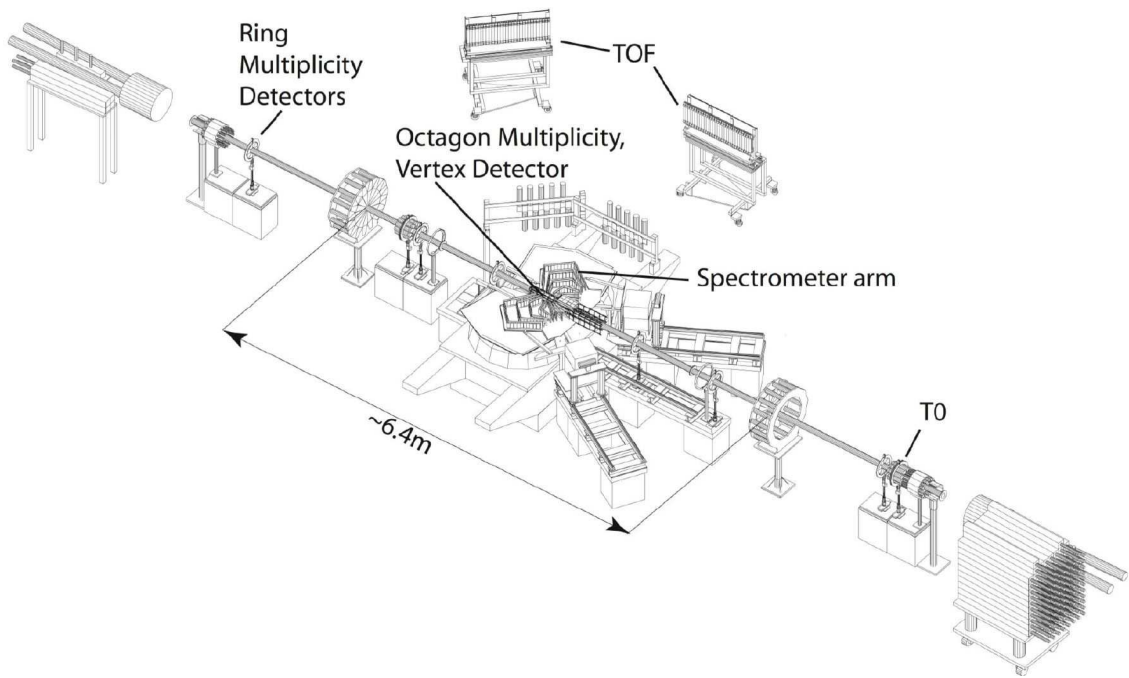,scale=1.0}
\caption{Layout of the PHOBOS experiment in 2003, the top part of the magnet around the collision point is removed for clarity.}
\label{Fig:phobos}
\end{center}
\end{figure*}

Its goal was to study particle production
down to very low $p_{T}$, 10~MeV/c for detectable and 30~MeV/c for momentum analyzed, over a large phase space. 
Four subsystems make up the experiment, see Fig.\ref{Fig:phobos}
a multiplicity array, a vertex detector, a two arm magnetic spectrometer including TOF, and several trigger detectors which also determine the 
centrality of the collision. Three of those consists of silicon pad arrays with a total 9 different silicon wafers with varying pad sizes. 
Details about these can be found in \cite{PHOBOS01} and information about tracking and vertex algorithms used in \cite{wozniak2005,kulinich2005}. After 
another very successful run PHOBOS was shutdown in 2005.\\

{\bf S}olenoidal {\bf T}racker {\bf a}t {\bf R}HIC (STAR) \cite{STAR01}. 
The collaboration consists of 616 physicists from 52 institutions in 12 countries, 
This experiment excels in measuring hadron production over a large solid angle. It features detector systems for high precision tracking, 
momentum analysis, and particle identification 
at center-of-mass rapidity. The centerpiece is the large TPC ($Ar/CH_{4}$ 90:10) with its two 210~cm drift regions ($|\eta|<1.8$) for tracking and PID 
together with a 3 layer silicon vertex tracker (SVT), a single layer silicon strip detector (SSD) 
and electromagnetic calorimetry around the TPC. Tracking of the around 2000 charged particles per central event 
in the TPC is performed by a Kalman Filter and Hough transform which points the tracks back to the SVT/SSD. 
Smaller angles, $\theta=2^{\circ}-9^{\circ}$, are covered by two 
small forward-TPCs ($Ar/CO_{2}$ 50:50, 23~cm drift, 240-1400 V/cm) with radial drift. 
Not only is the large volume of the TPC very sensitive to beam induced background, 
but the high collision rate at the large luminosities have introduced space charge problems as described in \cite{gene}. 
Over the years the SVT has suffered problems with a reset to its front-end electronics (FEE) crates which eliminated 
it from the data-stream for some fraction of events. \\

All experiments have performed beyond their expectations despite startup problems in the early runs. 
No beam induced aging has been observed in any of 
the tracking devices, most problems were related to the FEE. Small improvements and additions were made over the years, 
but more substantial upgrades are necessary to deal with future machine improvements.

\section{Upgrades}
\label{Sec:upgrades}

The quest to understand the new form of matter created in these nucleus collisions demand higher luminosities, 
which can be achieved by more numerous and smaller bunches, faster data acquisition (DAQ) rates, and 
new specialized detector systems. Beyond 2008 RHIC will also compete with the heavy ion program at the LHC and needs to exploit 
its unique spin physics program. Accelerator and detector upgrades are being proposed and built to keep RHIC and its 
experiments on the leading edge. A complete summary can be found in the reviews, white-papers and decadal plans in \cite{henp}.

\subsection{RHIC upgrades}
\label{SubSec:UpgrRHIC}

In the near future RHIC will upgrade its aging Tandem Van-de-Graaffs with an Electron Beam Ion Source (EBIS) \cite{ebis}. 
This source and linac based pre-injector can produce all ion species up to uranium, including noble gases and polarized $He^{3}$ 
and switch from one species to another within seconds to provide RHIC and the NASA Space Radiation Laboratory (NSRL) at BNL with a variety 
of ion species. EBIS can go into operation around 2009.

Currently intra-beam scattering limits the size and intensity of RHIC bunches. To achieve the planned upgrade to 40x design luminosity of RHIC-II
 electron cooling is needed. Electron cooling of a 100~GeV Au beam with about $10^{9}$ ions per bunch would require a 54~MeV electron beam 
of 100-200~mA current \cite{rhiccooling}. Prototype designs are underway for the required 26~m long, 2-5~T superconduction solenoid magnets, 
which require $10^{-5}$ field precision.

A high energy, high luminosity polarized electron-ion collider could study the fundamental structure of matter using deep inelastic scattering, 
RHIC provides already the ion part, so a eRHIC ZDR \cite{erhic} has been prepared which considers various aspects of the accelerator design.
The main design being a 5-10~GeV polarized electron (10~GeV positron) beam circulating in a storage ring which incepts the ion or 
polarized proton beam of RHIC in 
one interaction region. The expected luminosities for 10~GeV e on 250~GeV p are $\mathcal{L}=10^{32}-10^{33} cm^{-2}s^{-1}$ and 
$\mathcal{L}=10^{30}-10^{31} cm^{-2}s^{-1}$
for the 10~GeV e on 100~GeV ion beams. 70\% polarization, transverse and longitudinal, is expected for the lepton and proton beams.

\subsection{Experimental upgrades}
\label{SubSec:UpgrExp}

The RHIC baseline program will come to a close in a few years with the completion of PHOBOS (2005) and later BRAHMS. 
To enhance their detector capabilities over the coming years PHENIX and STAR have developed R\&D programs to cover the following   
main physics goals \cite{drees2004}:

\begin{itemize}
 \item Study of quark-gluon matter at high temperatures with AA, pA and pp collisions:
  \begin{itemize}
   \item Jet tomography, including identified hadrons in the $p_{T}$  range 3-10~GeV/c, hadron-jet and $\gamma-jet$ correlations, as well as flavor tagging.
   \item Thermal radiation and effects of chiral symmetry restoration in the electron-pair continuum, in particular at low masses $(<1~GeV/c^{2})$.
   \item Production of mesons with open charm and beauty.
   \item Quarkonia spectroscopy with $J/\psi$, $\psi'$, $\chi_{c}$, $\Upsilon(1s)$, $\Upsilon(2s)$, $\Upsilon(3s)$.
  \end{itemize}
 \item The extended exploration of the spin structure of the nucleon:
  \begin{itemize}
   \item Gluon spin structure $(\Delta G/G)$ with heavy flavour and $\gamma-jet$ correlations.
   \item Quark spin structure  $(\Delta q/q)$ with W-production.
   \item Transversity with jets.
  \end{itemize}
 \item Exploration of the nucleon structure in nuclei: A-, $p_{T}$-, x-dependence of the  parton structure of nuclei.
\end{itemize}

\begin{table*}[htdp]
\caption{ Upgrade plans for PHENIX and STAR and their relevance to the physics goals, X~-~upgrade critical for success, O~-~upgrade significantly enhances program}
\begin{center}
\begin{scriptsize}
\begin{tabular}{ | l | c |  c   c   c   c | c   c | c | } \hline

Upgrades & time  & \multicolumn{4}{c|}{High T QCD$  \ldots$   QGP} & \multicolumn{2}{c|}{Spin} & Low-x \\ \cline{3-8}

         & frame &              & heavy  & jet        &           &   &              &     \\ 
         &       & $e^{+}e^{-}$ & flavor & tomography & quarkonia & W & $\Delta G/G$ &     \\ \hline

PHENIX \cite{drees04}\\ \hline

Hadron blind detector \cite{ravinovich}  & 2006/07 &  X    &              &                &           &   &              &      \\
Vertex Tracker        \cite{heuser,ogilvie2004}& 2007/08 &  X    &      X       &       O        &     O     &   &      X       &   O  \\
Muon Trigger          \cite{barish2004}        & 2008    &       &              &                &     O     & X &              &      \\ 
Forward cal.          \cite{barish2004}        & 2009    &       &              &       O        &     O     & O &              &   X  \\ \hline

STAR \cite{hallman2004}\\ \hline

Time of Flight        \cite{majka2004}         & 2006/07  &       &      O       &       X        &     O     &   &              &      \\
heavy flavor tracker  \cite{schweda,wieman2004}& 2008     &       &      X       &                &     X     &   &              &      \\
Forward Tracker       \cite{hallman2004}       & 2010     &       &      O       &                &           & X &       O      &      \\
Forward Cal           \cite{hallman2004}       & 2006/07  &       &              &                &           &   &       O      &  X   \\
DAQ                   \cite{hallman2004}       & 2006/09  &       &      O       &       X        &     X     & O &       O      &  O   \\ \hline

RHIC Luminosity       \cite{satogata2004}      & 2009 -   &  O    &      O       &       X        &     X     & O &       O      &  O  \\ \hline

\end{tabular}
\end{scriptsize}
\end{center}
\label{Tab:upgrades}
\end{table*}

Table~\ref{Tab:upgrades} puts the upgrade plans into perspective with the physics goals they cover and indicates their time-frame and references.

\begin{figure*}[htbp]
\begin{center}
\epsfig{file=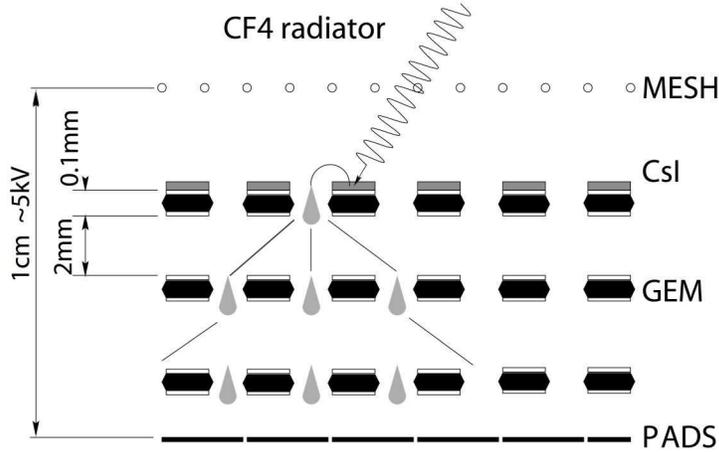,scale=1.0}
\caption{Proposed triple GEM readout for the PHENIX Hadron Blind Detector}
\label{Fig:hbdgem}
\end{center}
\end{figure*}

A Hadron Blind Detector (HBD) will be installed close to the collision point inside the PHENIX central magnet. 
Its a windowless Cherenkov detector with $CF_4$ as the radiator and chamber gas. 
The readout consists of a triple GEM chamber with a CsI coating on the top layer and pad read-out, see Fig.\ref{Fig:hbdgem}. The pad size is slightly smaller 
than the blob size expected from a passing hadron, resulting in a nearly 100\% probability for a single pad hit. 
Electrons are expected to produce about 35 photo-electrons which spread over at most 3 pads with the chance of a single-pad hit being negligibly small. 
The relatively large pad size results also in a low granularity and therefore
a low cost detector. In addition, since the photoelectrons produced by a single electron
will be distributed between at most three pads, one can expect a primary charge of at
least 10 electrons/pad, allowing operation of the detector at a relatively moderate gain
of a few times $10^3$.
Figure~\ref{Fig:hbdgem} illustrates the arrangement of the detector where the emitted electron is reflected by the electric field and amplified by the GEM layers which
in return shields the CsI layer from photons produced in the avalanche.

Both PHENIX and STAR plan to update their tracking and offset vertex reconstruction capabilities by new silicon vertex detectors. 
PHENIX foresees a 4 layer barrel section with pixels at r=2.5 and 5.0~cm and stripixels \cite{li2004} at r=10 and 14~cm, 
read out by the ALICE1LHCb and the FNAL SVX4 chip. 
In the forward direction 4 umbrella shaped silicon strip detectors will aid vertex reconstruction and tracking for the muon arms.

STAR has a two fold approach. Two layers of 30x30 $\mu m$ CMOS pixels at r=1.5 and 5.0~cm will reconstruct offset vertices from heavy flavor decays. 
In a second step a 4 layer silicon detector combined with forward GEM tracking chambers will further enhance the track reconstruction for small x.
 
To enhance their PID both experiments will install multi step resistive plate chambers (MRPC) as TOF detectors. 
STAR has several years experience with a small prototype and 
will complete the full barrel section in a few years.
PHENIX will install several modules of the same design in one of the central tracking arms. 
Single layer RPCs will also be used as trigger chambers for the PHENIX muon arms. 

In the very forward direction both experiments will install calorimeters for photon and $\pi^0$ detection. PHENIX plans to replace their Brass nose-cones with 
a Nose-Cone-Calorimeter (NCC) made from Tungsten plates interleaved with Si pad detectors. STAR will install an array of Pb-glass with photomultiplier readout inside the RHIC tunnel.

To cope with the high rates of the increased luminosity STAR will replace their TPC readout cards with the ALICE design over the next few years

\subsection{R2D}
\label{SubSec:R2D}

\begin{figure*}[htbp]
\begin{center}
\epsfig{file=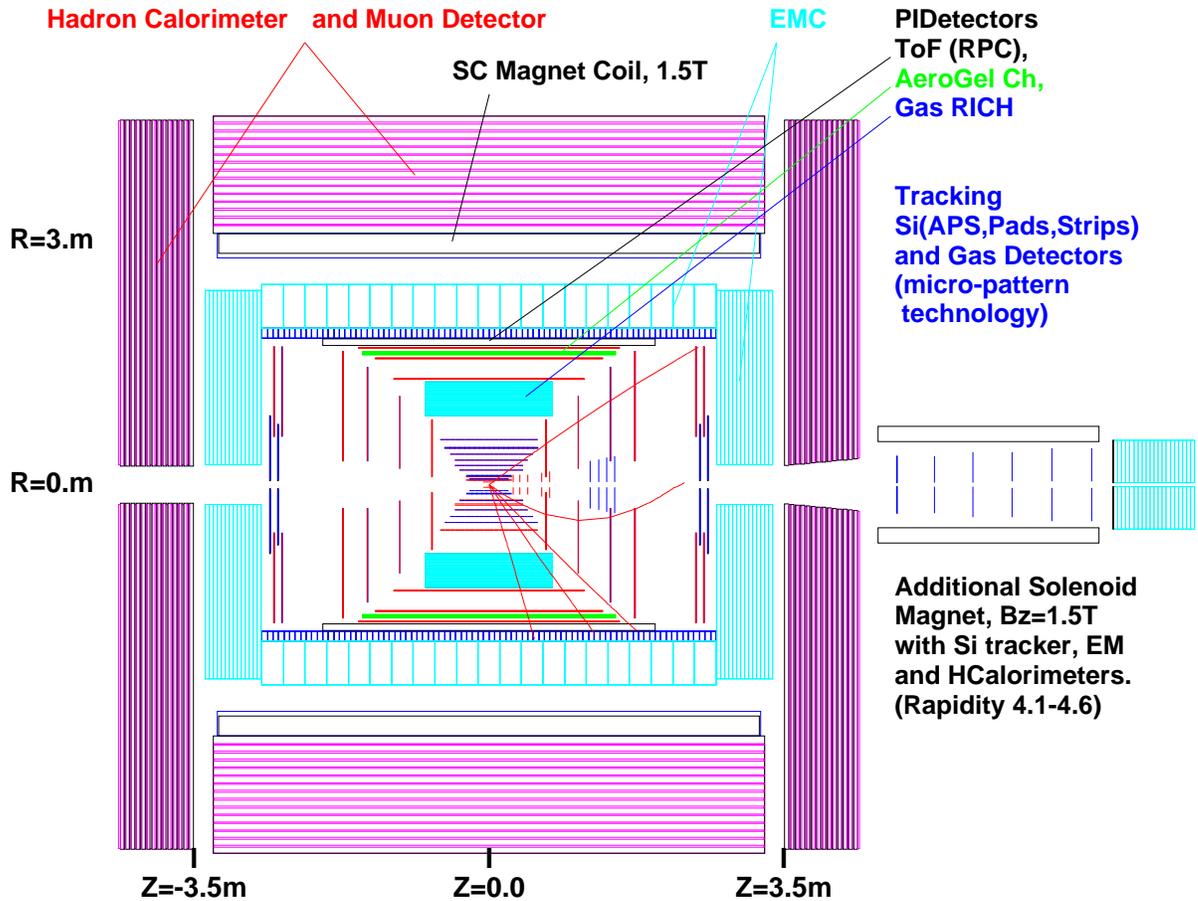,scale=1.0}
\caption{Diagram of a possible, comprehensive new detector
at RHIC II using the SLD magnet \cite{r2d}.}
\label{Fig:r2d}
\end{center}
\end{figure*}

To take full advantage of the luminosity available with RHIC II and to undertake detailed measurements of
jets, heavy flavors and electromagnetic probes over a large phase space $(-3~\le~\eta~\le~3, \Delta\phi~=~2\pi)$, 
a comprehensive new detector, see Fig.\ref{Fig:r2d}, has been proposed \cite{r2d}. Utilizing a high magnetic field (1.5~T) excellent 
momentum resolution is foreseen up to $p_{T}=40~GeV/c$ and the combination of calorimetry, dE/dx, TOF, Aerogel Threshold Cherenkov
and RICH counters will provide particle identification and lepton (e/h, $\pi$/h) separation from 1 up to 20~GeV/c in $p_{T}$. Even though this 
new detector would be superior to PHENIX and STAR separately, it does require a major investment. The physics reach beyond what can be achieved with the
moderate upgrades of PHENIX and STAR would be the measure to justify its construction.

\section{Summary and Acknowledgment}
\label{Sec:summary}

The first 5 years of RHIC operations have produced a wealth of data giving new insights into the complex world of 
high energy nuclear collisions and first glimpse at the role the gluon plays in the proton spin. More than a hundred publications 
and PhD thesis and two comprehensive summaries of the project itself and the physics of the first three years have been 
published. The accelerator has performed better than can be expected from a brand new machine and all the experiments have 
provided the measurements they were designed for. R\&D is ongoing at the experiments and accelerator to further push the 
luminosity and physics capability of the RHIC complex. In the more distant future eRHIC will open new physic domains to be studied. 

The author would like to thank all funding agencies, the staff of all experiments, and the 
collider group for their professional support to make RHIC such a great success.
                                   



\begin{thebibliography}{00}




\bibitem{baum2001}
G. Baum, Nucl.Phys. A698(2002) xxiii-xxxii

\bibitem{whitepaper}
'First Three Years of RHIC Operation, Physics perspectives from the experiments', Nucl. Phys A 757 (2005), 1-283 

\bibitem{rhicnim01}
H. Hahn et al., NIM A499 (2003) 245-263

\bibitem{rhicnim}
M. Harrison, T. Ludlum and S. Ozaki, NIM A499 (2003) 235-880

\bibitem{pakprivate} 
P. Steinberg, R. Pak, PHOBOS experiment, private communication.

\bibitem{rhiclumi}
'run overview of the relativistic heavy ion collider', http://www.agsrhichome.bnl.gov/RHIC/Runs/, assembled by W.Fischer, BNL.

\bibitem{EPAC04} 
9th European Particle Accelerator Conference, \newline
EPAC04, July 5-9, 2004, Lucerne, Switzerland, \newline
http://www.epac04.ch/

\bibitem{PAC05}
21st Particle Accelerator Conference, \newline
PAC05, May 16-20, 2005, Knoxville, TN, USA \newline
http://www.sns.gov/pac05/

\bibitem{eskola2001}
K.J. Eskola, Nucl.Phys. A698(2002) 78c-87c

\bibitem{Roland0510042}
'New Results from the PHOBOS Experiment', G. Roland, QM05 proceedings, arXiv:nucl-ex/0510042, Oct 2005

\bibitem{roland2005}
C. Roland, CMS Heavy Ion Program, these proceedings

\bibitem{zdc01}
A.J. Baltz, C. Chasman, S.N. White, NIM A417 (1989) 1-8

\bibitem{zdc02}
C. Adler et al., Nucl.Instrum.Meth. A470 (2001) 488-499

\bibitem{BRAHMS01}
M. Harrison, T.Ludlum and S.Ozaki, NIM A499 (2003) 437-468, http://www4.rcf.bnl.gov/brahms/WWW/

\bibitem{PHENIX01}
M. Harrison, T.Ludlum and S.Ozaki, NIM A499 (2003), 469-602, http://www.phenix.bnl.gov/

\bibitem{akikawa}
H. Akikaawa et al., NIM A499 (2003) 537-548 

\bibitem{murata}
J. Murata et al., NIM A500 (2003) 309-317

\bibitem{adcoxcentral}
K. Adcox et al., NIM A499 (2003) 489-507

\bibitem{allen}
M. Allen et al., NIM A499 (2003) 549-559

\bibitem{heuser}
J. Heuser, NIM A511 (2003) 210-214

\bibitem{PHOBOS01}
M. Harrison, T.Ludlum and S.Ozaki, NIM A499 (2003), 603-623, http://www.phobos.bnl.gov/

\bibitem{wozniak2005}
'Vertex reconstruction algorithms in the PHOBS experiment at RHIC', K. Wozniak these proceedings

\bibitem{kulinich2005}
'String banana template method for tracking in high-multiplicity environments with significant multiple scattering', P. Kulinich these proceedings

\bibitem{STAR01}
M. Harrison, T.Ludlum and S.Ozaki, NIM A499 (2003), 624-813, http://www.star.bnl.gov/

\bibitem{gene}
'Correcting for distortions due to Ionization in the STAR TPC', G. vanBuren these proceedings

\bibitem{ebis}
Progress on test EBIS and the Design of an EBIS-based RHIC Preinjector, J.G.Alessi et al., in \cite{PAC05}

\bibitem{rhiccooling}
Electron Cooling Dynamics for RHIC, A. V. Fedotov et al., AIP Conf. Proc. 773, 415 (2005)

\bibitem{erhic}
eRHIC Zeroth - Order Design Report, \newline 
BNL C-A/AP Note 142, 2004, \newline
www.agsrhichome.bnl.gov/eRHIC/eRHIC\_ZDR.htm \newline
eRHIC, A Future Electron-Ion Collider at BNL, V.Ptitsyn et al., MOPLT170 in \cite{EPAC04}

\bibitem{drees2004}
A. Drees, J. Phys. G:Nucl. Part.Phys. 30 (2004) S1109-S1112

\bibitem{henp}
High Energy and Nuclear Physics (HENP) \newline
Program Advisory Committee web page \newline
http://www.bnl.gov/henp/

\bibitem{nsac04}
NSAC Subcommittee on Relativistic Heavy Ions, \newline
June 2-6, 2004, BNL, Agenda web page \newline
http://nsac2004.bnl.gov/Subc\_PublicAgenda\_f.htm

\bibitem{drees04}
'PHENIX - Future Program', A. Drees in \cite{nsac04}

\bibitem{ravinovich}
'A Hadron Blind Detector for the PHENIX Experiment', I. Ravinovich et al., QM05 proceedings, arXiv:nucl-ex/0510024, Oct 2005

\bibitem{ogilvie2004}
'Silicon Vertex Tracker', C. Ogilvie in \cite{nsac04}

\bibitem{barish2004}
'Nose Cone Calorimeter, W boson trigger', K. Barish in \cite{nsac04}

\bibitem{hallman2004}
'STAR - Future Program', T. Hallman in \cite{nsac04}

\bibitem{majka2004}
'PHENIX - TOF Barrel Upgrade', R. Majka in \cite{nsac04}

\bibitem{wieman2004}
'Silicon MicroVertex Detector', H. Wieman  in \cite{nsac04}

\bibitem{schweda}
'A Heavy Flavor Tracker for STAR', K. Schweda, QM05 proceedings, arXiv:nucl-ex/0510003, Oct 2005

\bibitem{satogata2004}
'RHIC Performance and Plans towards higher Luminosity and higer Polarization', T. Sagotata in \cite{EPAC04}

\bibitem{li2004}
Z. Li et al., NIM A535 (2004) 404-409

\bibitem{r2d}
J.W. Harris et al., nucl-ex/0407021, 
P.Steinberg et al., nucl-ex/0503002
\end{thebibliography}
\end{document}